\documentclass[reprint,superscriptaddress,aps,prb,twocolumn,floatfix,nolinenumbers]{revtex4-2}
\usepackage[utf8]{inputenc}
\usepackage{textcomp} 
\usepackage{setspace}
\usepackage{amsmath}
\usepackage{breqn}
\usepackage{graphicx}
\usepackage{verbatim}
\usepackage{float}

\usepackage{amsfonts}
\usepackage{amssymb}
\usepackage{xcolor}
\usepackage{soul}
\usepackage{tabularx}
\setcitestyle{super}
\usepackage[colorlinks,linkcolor=blue,anchorcolor=blue,citecolor=blue,urlcolor=black]{hyperref}
\DeclareGraphicsExtensions{.pdf,.eps,.png,.jpg,.mps}
\usepackage{booktabs}
\usepackage{makecell}
\usepackage{threeparttable}\usepackage{lineno}
\usepackage{mhchem} 
\usepackage{gensymb}
\usepackage{upgreek}

\begin{document}

\title{High-coherence parallelization in integrated photonics}
\author{Xuguang Zhang$^{1,\dagger}$, Zixuan Zhou$^{1,\dagger}$, Yijun Guo$^{1,\dagger}$, Minxue Zhuang$^{1}$, Warren Jin$^{2}$, Bitao Shen$^{1}$, Yujun Chen$^{1}$, Jiahui Huang$^{1}$, Zihan Tao$^{1}$, Ming Jin$^{1}$, Ruixuan Chen$^{1}$, Zhangfeng Ge$^{5}$, Zhou Fang$^{4}$, Ning Zhang$^{4}$, Yadong Liu$^{4}$, Pengfei Cai$^{4}$, Weiwei Hu$^{1}$, Haowen Shu$^{1}$, Dong Pan$^{4}$, John E. Bowers$^{2,*}$, Xingjun Wang$^{1,3,5,*}$ \& Lin Chang$^{1,3,*}$\\
\vspace{3pt}
$^1$State Key Laboratory of Advanced Optical Communications System and Networks, School of Electronics, Peking University, Beijing, China.\\
$^2$Department of Electrical and Computer 
Engineering, University of California Santa Barbara, Santa Barbara, CA, USA.\\
$^3$Frontiers Science Center for Nano-optoelectronics, Peking University, Beijing, China.\\
$^4$SiFotonics Technologies Co., Ltd., Beijing, China.\\
$^5$Peking University Yangtze Delta Institute of Optoelectronics, Nantong, China.\\
$^\dagger$These authors contributed equally to this work. \\
\vspace{3pt}
\centering{\small Corresponding authors: $^*$bowers@ece.ucsb.edu, $^*$xjwang@pku.edu.cn, $^*$linchang@pku.edu.cn}}

\maketitle
\noindent
\textbf{Abstract}\\ 
\noindent 
\textbf{Coherent optics has profoundly impacted diverse applications ranging from communications, LiDAR to quantum computations. However, building coherent systems in integrated photonics previously came at great expense in hardware integration and energy efficiency: the lack of a power-efficient way to generate highly coherent light necessitates bulky lasers and amplifiers, while frequency and phase recovery schemes require huge digital signal processing resources. In this work, we demonstrate a high-coherence parallelization strategy that facilitates advanced integrated coherent systems at a minimum price. Using a self-injection locked microcomb to injection lock a distributed feedback laser array, we boost the microcomb power by a record high gain of up to 60 dB on chip with no degradation in coherence. This strategy enables tens of highly coherent channels with an intrinsic linewidth down to the 10 Hz level and power of more than 20 dBm. The overall electrical to optical wall-plug efficiency reaches 19\%, comparable with that of the state-of-the-art semiconductor lasers. Driven by this parallel source, we demonstrate a silicon photonic communication link with an unprecedented data rate beyond 60 Tbit/s. Importantly, the high coherence we achieve reduces the coherent-related DSP consumption by 99.999\% compared with the traditional III-V laser pump scheme. This work paves a way to realizing scalable, high-performance coherent integrated photonic systems, potentially benefiting numerous applications.}\\


\begin{figure*}[ht]
    \centering
    \setlength{\abovecaptionskip}{-0.25cm}
    \includegraphics[width=\linewidth]{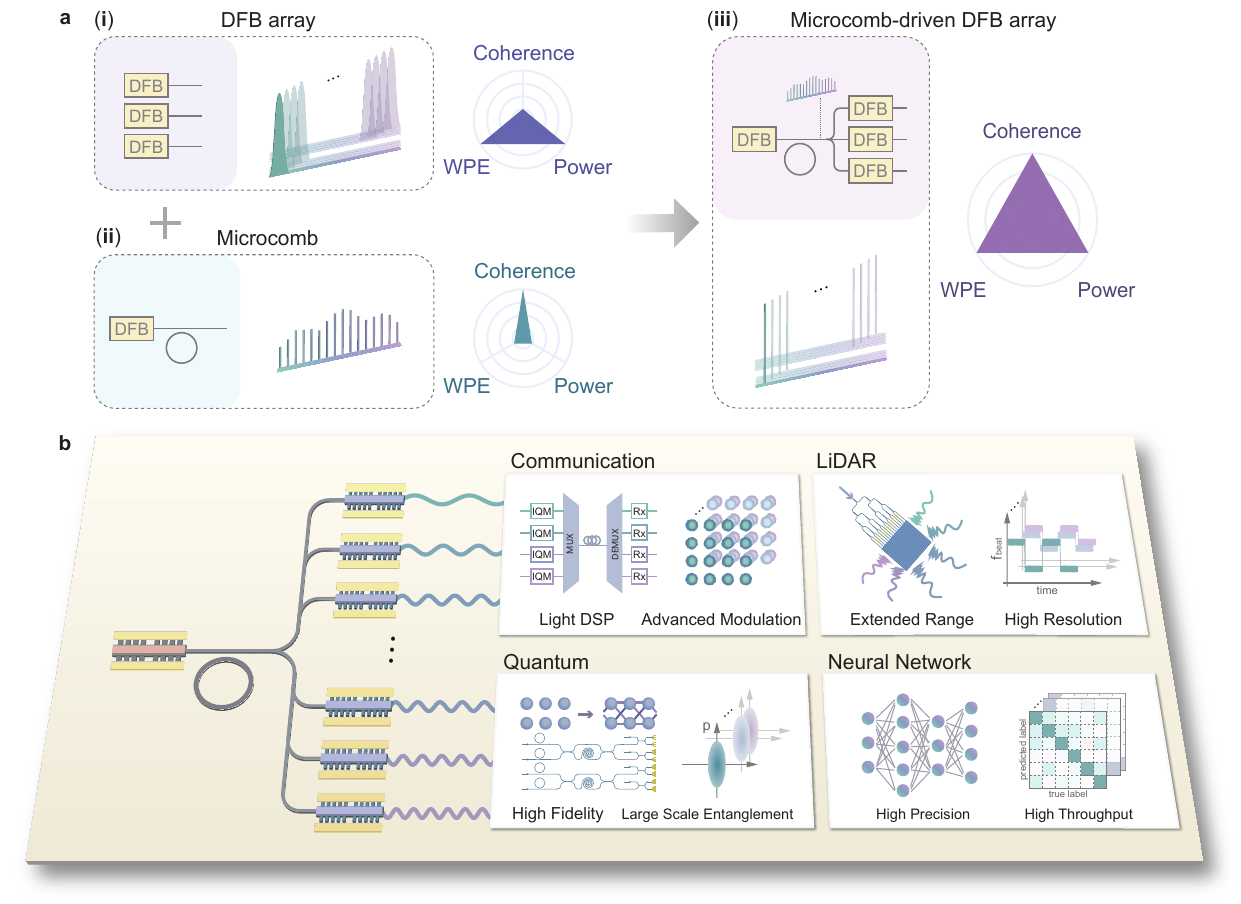}
    \caption{\textbf{High-coherence parallelization based systems.} 
    \textbf{a,} Using a self-injection locked DFB laser, we pump a Si$_3$N$_4$ microresonator to generate the microcomb, which is then employed to drive the DFB array (\textbf{iii}). We can obtain a highly coherent parallel light source that combines the advantages of high power and high WPE of the DFB array (\textbf{i}) and narrow linewidth of the microcomb (\textbf{ii}). \textbf{b,} Conceptual illustration of the high-coherence parallelization system. This light source can find wide applications in integrated photonic scenarios and exhibits significant potential across various fields, such as communications, LiDAR, quantum information and neural networks.} 
    \label{fig1}
    \vspace{-5pt}
\end{figure*}

\noindent\textbf{Introduction}\\
Integrated photonics has made remarkable progress in providing scalable solutions for communications \cite{pfeifle2014coherent,marin2017microresonator,yang2022multi,shu2022microcomb,rizzo2023massively}, computations \cite{shen2017deep,feldmann2021parallel,xu202111,ashtiani2022chip}, quantum information\cite{wang2020integrated, pelucchi2022potential} and sensing \cite{riemensberger2020massively,chen2023breaking,li2023frequency}. To meet the ever-growing capacity and precision requirements posed by these applications, one major evolution trend over the last few years is the adoption of coherent technology in photonic integrated circuits (PICs)\cite{dong2014monolithic,martin2018photonic}. This technology allows for the manipulation of frequency and phase, offering enormous application prospects: in communications, the data rate of a transceiver can be lifted by orders of magnitude when replacing the intensity modulation with a coherent scheme \cite{fulop2018high,corcoran2020ultra,lundberg2020phase,xu2022dual,jorgensen2022petabit}; in metrology, high optical coherence can enable frequency synthesizing and time-keeping with precision beyond $10^{-15}$, introducing atomic-level accuracy into PICs \cite{spencer2018optical}; in ranging, coherence detection supports FMCW LiDAR that can simultaneous capture the distance and speed information\cite{riemensberger2020massively}.\\
\indent However, despite great potential, building a coherent system in integrated photonics has come at a great price in hardware and power consumption. One major difficulty lies in the sources: so far, there is no approach that can generate light with high parallelism, high coherence, and high power simultaneously.
The most commonly used source on chip, the III-V distributed feedback (DFB) laser, has excellent power\cite{8672593} and electrical to optical wall-plug efficiency (WPE)\cite{vaskasi2022high}, but its intrinsic linewidth typically resides at the 100 kHz level, falling short of the coherence requirements in many applications. To achieve better coherence, a commonly adopted method is coupling III-V lasers to high-quality (Q) cavities, which can effectively reduce the linewidth to the sub-kHz level. Nevertheless, this sacrifices power and WPE\cite{shen2020integrated,xiang2021laser}. The high-Q resonator is also used to generate optical frequency combs in parallel coherent systems, but it further worsens both the power and WPE significantly: the nonlinear frequency conversion process usually has an optical-to-optical conversion efficiency at the order of only a few percent\cite{marin2017microresonator,stern2018battery}. Various methods have recently been developed to improve the conversion efficiency\cite{xue2017microresonator,li2022efficiency,helgason2022power}, but the fact that all the comb lines split energies from the pump ultimately limits the channel power, usually below -10 dBm. 
To leverage such low power in systems, strong amplification with a gain of more than 30 dB is often necessary, posing a challenge for both the integrated erbium-doped fiber amplifier (EDFA)\cite{shu2022microcomb,riemensberger2020massively} and the semiconductor optical amplifier (SOA)\cite{raja2021ultrafast}. Meanwhile, these two methods inevitably introduce noise into the amplification process.
As a result, at the system level, the coherent source still has to rely on expensive benchtop equipment, which is not suitable for scalable applications.\\
\indent Another challenge for integrated coherent systems is the huge digital signal processing (DSP) consumption\cite{pillai2014end,faruk2017digital}. Since the frequency and phase information need to be precisely recovered from the coherent detection, much more DSP is thus required over the intensity-based system. This requirement dramatically increases the power budget and often necessitates dedicated electronics such as DSP chips\cite{crivelli2013architecture}. 
To deploy advanced coherent communication in the next generation data center, DSP chips need to be manufactured with a 3-nm complementary metal-oxide-semiconductor (CMOS) process to alleviate high power consumption\cite{tauber2023role}. Furthermore, the complexity of the DSP makes real-time information processing difficult, as in the LiDAR application.
Several attempts have been made to reduce DSP, including laser synchronization\cite{brodnik2021optically} and cloned combs\cite{geng2022coherent,zhang2023clone}, but these approaches involve bulky narrow linewidth sources and desktop phase-locking loops, dramatically adding the hardware burdens at the system level.\\
\indent In this work, we overcome previous challenges by demonstrating a high-coherence parallelization strategy in integrated photonics. Instead of pushing the efficiency of the coherent light source itself, here we use a high-coherence microcomb as a seed to injection lock (IL) a DFB array (Fig. \ref{fig1}). We theoretically prove and experimentally demonstrate that this approach can combine the advantages of both the microcomb and the DFB, achieving high coherence, high power, and high WPE simultaneously. We show that this strategy can significantly boost the performance of integrated coherent systems by demonstrating a parallel coherent communication experiment using parallel silicon photonic (SiPh) transceivers. We achieve an unprecedented data rate of over 60 Tbit/s and nearly no extra coherent DSP consumption.\\ 

\begin{figure*}[ht]
    \centering
    \setlength{\abovecaptionskip}{-0.5pt}
    \includegraphics[width=\linewidth]{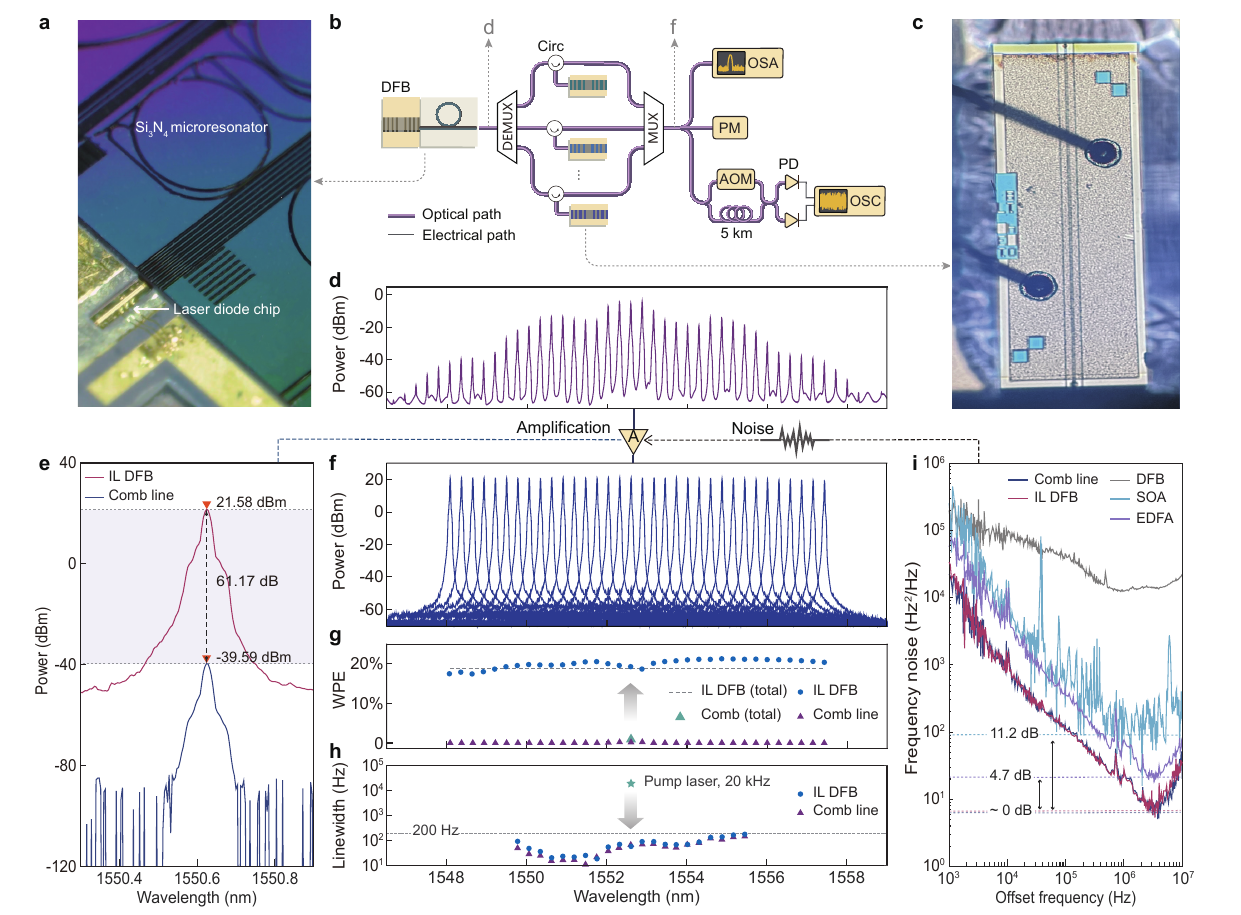}
        \caption{\textbf{Fundamental characteristics of the self-injection locked microcomb and injection locking amplification.} \textbf{a,} Photograph of the integrated DFB laser and a Si$_3$N$_4$ microresonator. \textbf{b,} The link used for fundamental characteristics measurement. DFB, distributed feedback laser; DEMUX, demultiplexer; Circ, circulator; MUX, multiplexer; OSA, optical spectrum analyzer; PM, power meter; AOM, acousto-optic modulator; PD, photodetector; OSC, oscilloscope. \textbf{c,} Photograph of the slave DFB laser. \textbf{d,} Spectrum of the microcomb pumped by the self-injection locked DFB laser. \textbf{e,} Spectra of the comb line at 1550.62 nm before and after IL amplification, show that a gain of more than 60 dB can be achieved. \textbf{f,} Combined spectra of each comb line after being filtered and IL amplified. \textbf{g,} WPE of different cases. The purple triangles and blue circles represent the WPE of each comb line before and after IL amplification. The green triangle represents the overall WPE of the original microcomb. The dash line represents the overall WPE of all comb lines after IL amplification. \textbf{h,} The purple triangles and blue circles represent the intrinsic linewidth of the comb lines before and after IL amplification, respectively. Both of them have linewidths below 200 Hz, while the DFB laser has a linewidth over 20 kHz. \textbf{i,} Single-sideband (SSB) frequency noise of different amplification methods. IL amplification has no significant noise-floor increment. }
    \label{fig2}
    \vspace{-5pt}
\end{figure*}
\noindent\textbf{Results}\\
\noindent\textbf{Power-efficient, high-coherence parallel light source on chip}\\
\noindent Generating the parallel light source consists of two steps. Firstly, we self-injection lock (SIL) an indium phosphide (InP) DFB diode laser to a 35 GHz free spectral range (FSR) Si$_3$N$_4$ microresonator with a thickness of 100 nm (Fig. \ref{fig2}a). The Rayleigh backscattering from the microresonator imparts a narrow spectral feedback to the DFB laser and compels the laser to oscillate at the resonance of the microresonator, leading to a strong frequency noise reduction of the laser diode\cite{shen2020integrated,jin2021hertz}. In this work, the microresonator exhibits a high intrinsic Q factor around 4.3 × $10^7$, which reduces the original 20 kHz intrinsic linewidth of the pump DFB to a linewidth down to the 10 Hz level \cite{camatel2008narrow,yuan2022correlated}.\\
\indent The SIL scheme also enables optical frequency comb generation\cite{chang2022integrated}. Since the microresonator is designed for normal dispersion, the microcomb operates in the dark pulse regime\cite{xue2015mode,shu2022microcomb}, ensuring good coherence over comb lines and distributing narrow-linewidth characteristics among all wavelengths. Our microcomb is pumped by a DFB laser with a power of 19 dBm, and the spectrum of the microcomb is depicted in Fig. \ref{fig2}d. The comb lines exhibit an intrinsic linewidth as low as 11 Hz (Fig. \ref{fig2}h). Our microcomb is among the state-of-the-art integrated optical frequency combs regarding coherency. The microcomb generation is turnkey \cite{shen2020integrated} without any auxiliary electronics control, and the comb state can stably last for several hours (Supplementary Note \uppercase\expandafter{\romannumeral 3} and \uppercase\expandafter{\romannumeral 4}).\\
\indent Secondly, the microcomb serves as a coherence seed to IL a DFB array\cite{heffernan202360} (Fig. \ref{fig2}c). We sequentially filter each comb line (master) and inject it into a DFB laser (slave), thereby locking the frequency of the slave laser (Fig. \ref{fig2}b and Methods). 
One advantage of this IL amplification is its strong power gain and uniform output power. 
The amplification process is described by the following equation\cite{lang1982injection, lau2009enhanced, liu2020optical} (Supplementary Note \uppercase\expandafter{\romannumeral 2}):
\begin{equation}
    S = \frac{ S_{fr}- \frac{\gamma_N}{\gamma_p} \Delta N}{1+\frac{g}{\gamma_p} \Delta N}
\end{equation}
where $\gamma_N$, $\gamma_P$, $g$ refer to the slave laser's carrier recombination rate, photon decay rate, and linear gain coefficient, respectively. $\Delta N$ denotes the carrier number difference between the injection-locked case and the free-running case. $S$ denotes the intracavity photon number, which is proportional to the output power $P$. The equation above describes the relationship between injection-locked and free-running intracavity photon numbers ($S$ and $S_{fr}$).\\
\indent Given the amplification circumstances where $\Delta N << 1$, the intracavity photon number closely approximates to the free-running case, $S \sim S_{fr}$. This indicates that the output power of the slave laser $P$ is nearly the same as the DFB free-running power $P_{fr}$ regardless of the original master light power. This significant amplification effect comes from the slave laser's cavity enhancement. IL-based amplification harnesses the media gain more effectively and thus leads to higher WPE compared with single-pass amplifiers such as EDFA and SOA.\\
\indent To verify this, we attenuate the power of a comb line to a significantly low level before injecting it into a DFB laser, whose free-running power can reach around 21 dBm. The experimental results show that even under the injected power of -39.6 dBm, the slave DFB laser can still be effectively locked and emits a power of 21.6 dBm, indicating an exceptional gain exceeding 61 dB (Fig. \ref{fig2}e). According to our knowledge, such gain value is 30 dB higher than those of other integrated amplifiers\cite{sobhanan2022semiconductor,liu2022photonic}. By applying this strategy to each channel of the SIL microcomb and combining them, we obtain a flat spectrum profile (Fig. \ref{fig2}f), with all 34 channels exhibiting a power exceeding 20 dBm. The WPE of each comb line after amplification is nearly the same as the free-running DFB, which is significantly higher than the original WPE of each comb line (Fig. \ref{fig2}g and Supplementary Note \uppercase\expandafter{\romannumeral 5}). The overall WPE, which includes the power consumption of the pump DFB for microcomb generation, is beyond 18.8\%. This WPE is almost independent of the original microcomb WPE (1\%), since the number of parallel channels is relatively large.\\
\indent Another appealing property of the IL process is the maintenance of high coherence from the seed light. 
The phase noise after IL is described by the following equation\cite{lau2009enhanced}(Supplementary Note \uppercase\expandafter{\romannumeral 2}): 
\begin{equation}
\begin{aligned}
    <\delta \phi \delta \phi>_\omega = |H_{\phi,\phi_{inj}}(\omega)|^2 <\delta \phi_{inj} \delta \phi_{inj}>_\omega \\
    + |H_{\phi,\phi}(\omega)|^2 <F_{\phi} F_{\phi}>_\omega + \Delta
\end{aligned}
\end{equation}
where $<\delta \phi_{inj}\delta \phi_{inj}>_\omega$ and $<F_{\phi}F_{\phi}>_\omega$ represent 
the master light phase fluctuation power spectrum density (PSD) and the slave laser cavity Langevin noise in phase, while $\Delta$ represents the phase and photon/carrier number coupling terms that are relatively insignificant in the IL process.
Specifically, the frequency response of laser cavity 
Langevin noise $|H_{\phi,\phi}(\omega)|^2$ can be suppressed by more than 40 dB compared with the free-running case, while that of the master light $|H_{\phi,\phi_{inj}}|$ remains near 1 in the frequency range of our interest. 
In other words, the noise factor of the IL process is $\sim 0$ dB.\\
\indent Here we compare the degradation of noise under various amplification strategies. We use a comb line with a power of -15.4 dBm and amplify it by 20 dB gain through EDFA, SOA and injection locked DFB respectively. 
Figure \ref{fig2}i displays the single-sideband (SSB) frequency noise spectra of the input and amplified light. It is observed that incorporating EDFA and SOA has noise-floor increments of 4.7 dB and 11.2 dB, respectively, whereas the IL amplification shows no appreciable increase in the noise-floor. Compared with the EDFA/SOA case in which the amplification effect accumulates along propagation, the light field circulates and gets enhanced in the slave laser cavity. Consequently, more carriers are involved in stimulated emission rather than spontaneous emission, resulting in reduced additional amplifier spontaneous emission (ASE) noise and better phase noise performance. We further measure the linewidth of the IL amplified comb lines, each of which is at an output power of 20 dBm. It is shown that the coherence of the original microcomb is preserved over the spectrum (Fig. \ref{fig2}h and methods).\\

\begin{figure*}[ht]
    \centering
    \setlength{\abovecaptionskip}{-0.5pt}
    \includegraphics[width=\linewidth]{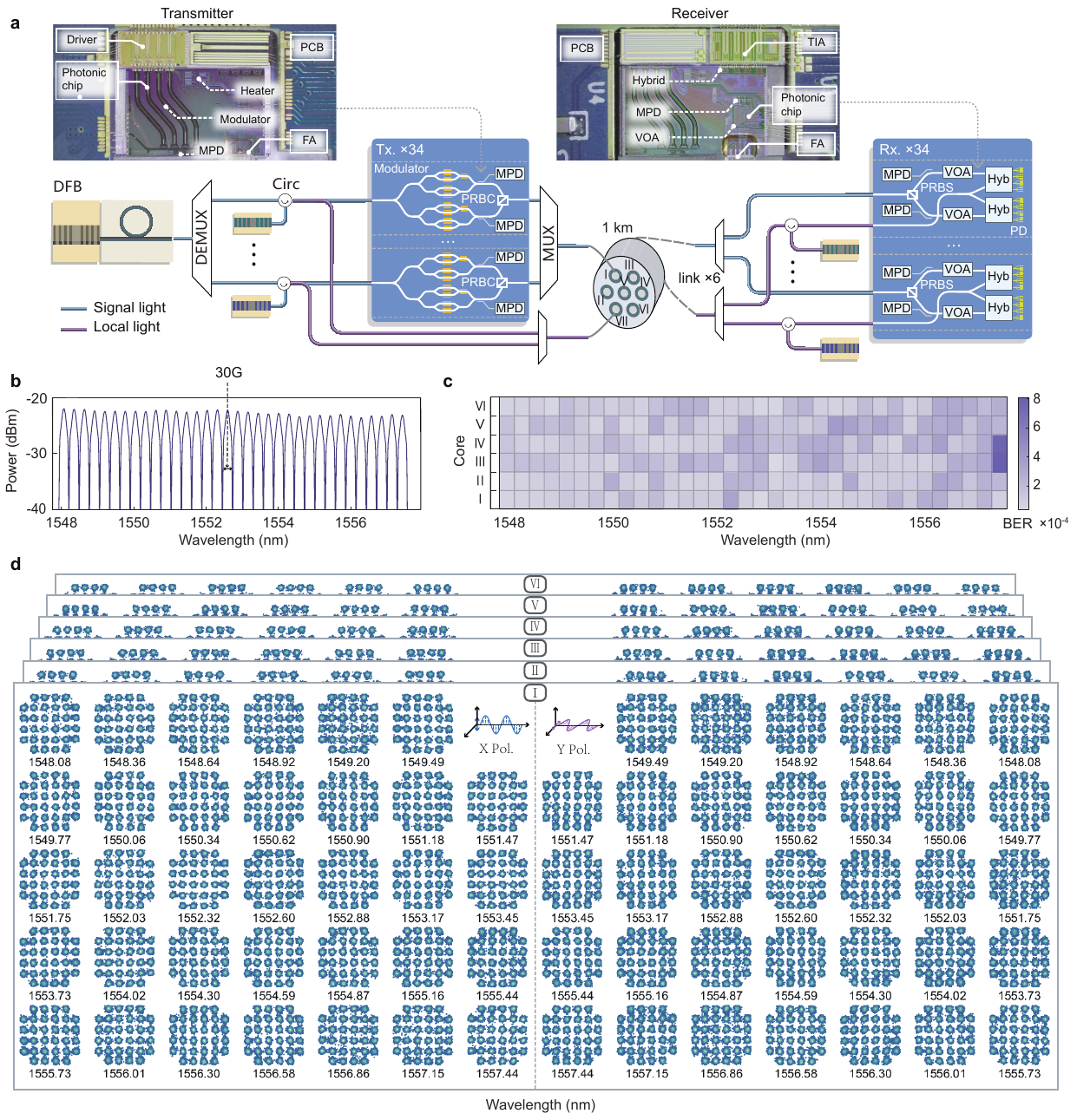}
    \caption{\textbf{Parallel coherent silicon photonic communications.} \textbf{a,} Architecture of the coherent communication system utilizing our power-efficient, high-coherence parallel source. Photographs of the transmitter and receiver chips are shown at the top. Tx, transmitter; Rx, receiver; PRBC, polarization rotator and beam combiner; PRBS, polarization rotator and beam splitter; MPD, monitor photodetector; VOA, variable optical attenuator; Hyb, $90^\circ$ hybrid; FA, fiber array; TIA, trans-impedance amplifier. \textbf{b,} Combined spectra of modulated carriers for all 34 channels. \textbf{c,} BERs across 34 wavelengths and 6 cores. \textbf{d,} Recovered constellation diagrams across 34 channels and 6 cores. X Pol, X polarization; Y Pol, Y polarization.}
    \label{fig3}
    \vspace{-5pt}
\end{figure*}
\noindent\textbf{High-capacity parallel communications with silicon photonic tranceivers}\\
\noindent This power-efficient, high-coherence parallel source can significantly enhance the performance of coherent systems on PICs and enable complex system functionality. 
Here, we use it to drive SiPh transceivers in a high-speed coherent optical communication system. Figure \ref{fig3}a illustrates the system architecture. 
The transmitter necessitates a power of more than 15 dBm to accommodate advanced modulation formats. 
To meet this requirement, each comb line is separated into different channels by a demultiplexer (DEMUX) and amplified to a uniform power of about 20 dBm by injection locking the DFB, with an optical carrier-to-noise-ratio (OCNR) exceeding 70 dB.
Within each channel, the light is divided into two parts. 
One, serving as a data carrier, holds 99\% of the power and undergoes modulation by a SiPh transmitter, while the other remains unmodulated, retaining only 1\% of the power, and functions as the local oscillator (LO). 
The modulated data channels are combined by a multiplexer (MUX), and so are the LOs. 
To ensure high coherence for homodyne detection, the data carriers and the LOs are transmitted concurrently. 
At the receiver side, the combined data carriers and LOs are separated into different wavelength channels, each of which is directed to the corresponding receiver. 
To satisfy the power requirements of receivers and further enhance detection accuracy, the LOs are boosted to approximately 20 dBm by injection locking other DFBs. 
Thanks to the ample power and excellent coherence, our approach can effectively facilitate wavelength division multiplexing (WDM) coherent transmission of high capacity and advanced modulation formats.\\
\indent It's noteworthy that both the transmitter and the receiver utilized in our experimental setup are silicon-on-insulator (SOI) based and manufactured by a commercial CMOS foundry, as depicted in Fig. \ref{fig3}a. The transmitter comprises two IQ modulators, one polarization rotator and beam combiner (PRBC), two monitor photodetectors (MPD) and associated drivers. 
The receiver includes dual $90^\circ$ optical hybrids, one polarization rotator and beam splitter (PRBS), two MPDs, variable optical attenuators (VOA), four high-speed balanced-photodiodes (BPD), and trans-impedance amplifiers (TIA). 
We have optically and electrically packaged the SiPh transmitter and receiver, which possess electro-optic bandwidth and opto-electrical bandwidth of around 50 GHz, respectively.\\
\indent We employ dual-polarization 32-state quadrature amplitude modulation (DP-32-QAM) at a symbol rate of 30 Gbaud to encode data on each of the 34 amplified comb lines. 
Consequently, the data rate of each modulated channel is 300 Gbit/s (30 GBaud $\times$ 5 bits $\times$ 2 polarizations). 
These 34 channels are aggregated via a MUX, resulting in an overall data rate of 10.2 Tbit/s.
Moreover, we utilize a 1-km-long 7-core fiber, with 6 cores allocated for transmitting the modulated light and 1 core dedicated to transmitting the LOs at minimal power levels. 
After signal recovery at the receiver side, the calculated bit error ratios (BER) of all channels remain below $8.1\times10^{-4}$ (Fig. \ref{fig3}c), much lower than the 7\% hard-decision forward error correction (HD-FEC) threshold of $3.8\times10^{-3}$. Figure \ref{fig3}d displays the constellation diagrams of recovered dual-polarization data for all channels in cores 1-6 after transmission(Supplementary Note \uppercase\expandafter{\romannumeral 6}). 
The clear constellation diagrams of all channels demonstrate the excellent behavior of our approach in advanced coherent modulated formats.
Finally, we validate a silicon-based coherent WDM transmission experiment with an aggregate gross line rate of 61.2 Tbit/s and a net spectral efficiency of 8.01 bit/s/Hz (Fig. \ref{fig3}b).
According to our knowledge, this is the record number for a coherent link based on integrated sources and SiPh transceivers, marking a rate 30 times higher than that of the existing SiPh links\cite{shu2022microcomb}.\\ 

\begin{figure*}[ht]
    \centering
    \setlength{\abovecaptionskip}{-0.5pt}
    \includegraphics[width=\linewidth]{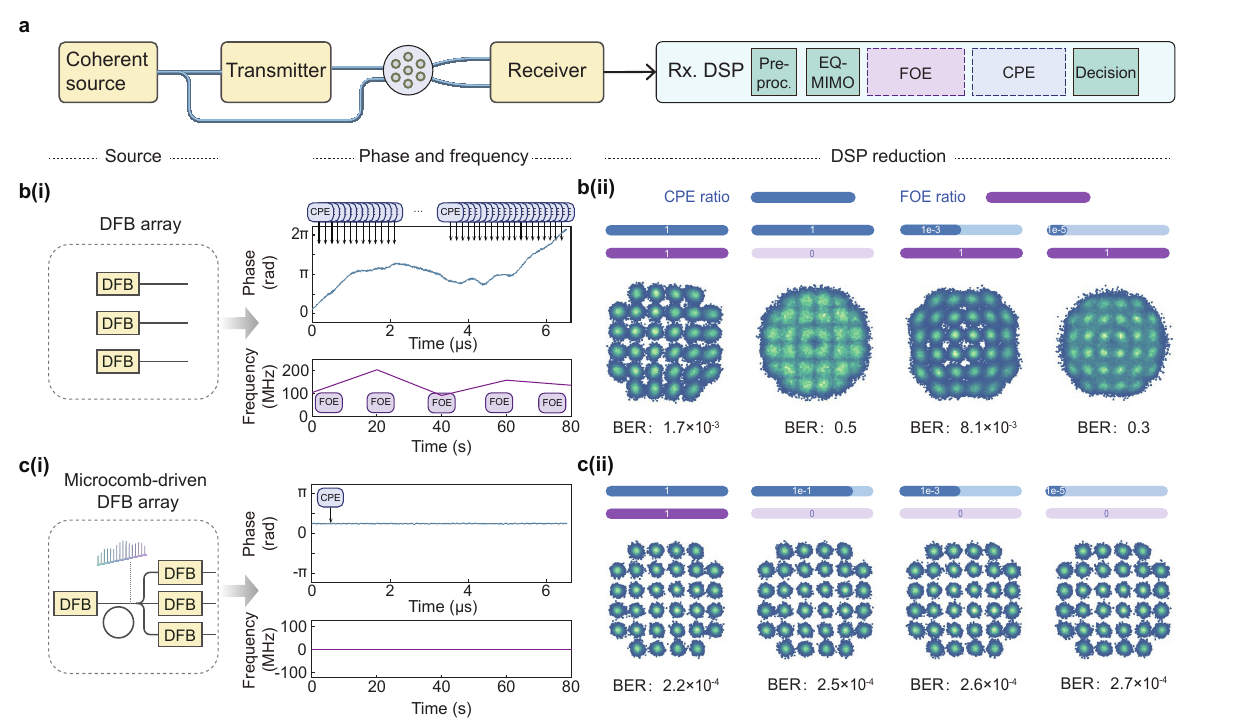}
    \caption{\textbf{DSP reduction induced by high-coherence parallelization. a,} Brief architecture and DSP modules in our strategy. Pre-proc, pre-processing; EQ-MIMO, equalization multiple-input-multiple-output; FOE, frequency offset estimation; CPE, carrier phase estimation. \textbf{b(i),} Using standalone DFBs as the carrier and the LO, the frequency offset and the phase difference change rapidly, requiring frequent CPE and FOE in DSP. 
    \textbf{b(ii),} The application of various DSP reduction results in the degradation of constellation diagrams.
    \textbf{c(i),} In our scheme, the frequency offset is 0, and the phase difference is stable, thus significantly reducing the ratio of CPE and omitting FOE. 
    \textbf{c(ii),} Recovered constellation diagrams with different ratios of CPE and FOE when using our strategy.}
    \label{fig4}
    \vspace{-5pt}
\end{figure*}
\noindent\textbf{DSP reduction in integrated high-coherence parallel system}\\
\noindent Besides the hardware breakthrough, this parallel approach can also bring significant DSP reduction in coherent systems. Conventional coherent communication systems usually use DFBs to generate optical carriers and LOs, whose frequencies and phases vary rapidly and randomly. Consequently, unpredictable frequency offsets and phase fluctuations arise between carriers and LOs, as depicted in Fig. \ref{fig4}b(i). To accurately recover the signal, frequency offset estimation (FOE) and carrier phase estimation (CPE) become necessary. 
This brings additional DSP requirements for coherent communications compared to intensity modulation direct detection (IM-DD) systems, consequently leading to higher power consumption.\\
\indent There are two key advantages of our strategy for saving the DSP. Firstly, the source itself exhibits a narrow linewidth. Secondly, our strategy preserves coherence during amplification, allowing for the low-power propagation of the LO signal.
The coherent communication link and DSP modules we use are illustrated in Fig. \ref{fig4}a. 
In our strategy, we ensure the absence of frequency offset and minimize phase fluctuations between the signal carrier and the LO, as illustrated in Fig. \ref{fig4}c(i). 
Consequently, FOE can be entirely omitted. 
Moreover, by simply estimating the initial phase and applying it to subsequent data, we can still maintain relatively high performance.\\
\indent To demonstrate the impact of reducing the usage of coherent DSP, we apply different ratios of FOE and CPE to a 30 GBaud 32-QAM coherent transmission link (Methods). We also compare the data recovery accuracies between our strategy and using two standalone DFB lasers as the carrier and the LO. 
Both schemes recover the signal successfully when employing full coherent-related DSP which incorporates FOE once and CPE for every symbol. 
Utilizing two standalone DFB lasers results in a BER of $1.7\times10^{-3}$, whereas our strategy achieves a significantly lower BER of $2.2\times10^{-4}$.
Furthermore, in the case of using standalone DFB lasers without FOE, recovery becomes infeasible due to the significant variation in frequency offset. 
When we decrease the normalized ratio of CPE while incorporating FOE, a CPE ratio of $1 \times 10^{-3}$ results in a BER of $8.1 \times 10^{-3}$. 
As the CPE ratio continues to decrease, data recovery becomes impossible (Fig. \ref{fig4}b(ii)). 
In our scheme, we benefit from the complete removal of FOE and a substantial reduction in the CPE ratio. The CPE ratio represents the number of CPE operations we perform compared to the traditional algorithm, which executes a CPE operation on every symbol. We gradually reduce the CPE ratio to $1\times10^{-5}$, meaning we run CPE once every other $1\times10^{5}$ data blocks, and still achieve successful data recovery, demonstrating a BER performance of $2.7\times10^{-4}$, which is comparable to a full DSP (Fig. \ref{fig4}c(ii)). 
Notably, the minimum required CPE ratio is currently limited by the length of our data within a time window. 
The comparison between the two schemes highlights the strong performance of our strategy, capable of reducing 99.999\% of the coherent-related DSP (FOE and CPE).\\

\noindent\textbf{Discussion}\\
\noindent By utilizing microcombs with much more channels and broader spans as we demonstrated before\cite{briles2021hybrid}, the parallelization level can be significantly improved. In our experiment, we employ a filter to select each comb line for the WDM purpose, which can later be substituted by other integrated components, such as arrayed waveguide gratings. 
It's noteworthy that for injection locking, the circulators or isolators might not be necessary. 
Employing a power splitter or injecting light through the second port of the laser can enable the same function and is more integration-friendly.\\
\indent Beyond delivering exceptional performance, our strategy offers compatibility with the existing photonic ecosystem.
The III-V laser array has been extensively used in photonic transceivers\cite{yang2023single}, LiDAR\cite{dilazaro2017multi} and computation units\cite{nahmias2018teramac}, and the transition to our highly-coherent source only takes one chip-based comb generator. Importantly, all the key components in our systems are chip-based and manufactured from photonic foundries. Our recent advancements in heterogeneous fabrication technology potentially can enable the integration of all these components on the same substrate, supporting mass production\cite{xiang2021laser,liang2022energy,tran2022extending}. The simplification of the coherent DSP, together with the emergence of the linear-drive pluggable optics (LPO) technologies, paves the way for a DSP-light coherent system with great energy efficiency.\\
\indent Our method can be extended to diverse applications beyond communications. For FMCW LiDAR, achieving more than 20-dBm power at multiple channels with high coherence is critical for long-distance ranging in automobiles with a fast scan rate\cite{zhang2022demonstration}. 
The high-power, high-coherence parallelization can also improve the signal-to-noise ratio in computation PICs, particularly when thousands of cascaded MZIs introduce large losses\cite{shen2017deep}. 
We also expect this strategy to facilitate the f-2f self-reference process\cite{del2016phase,okawachi2020chip,briles2021hybrid} by boosting the power for nonlinear frequency conversions, which will benefit optical synthesizers\cite{spencer2018optical} and optical clocks\cite{newman2019architecture} on chip.\\

\bibliographystyle{naturemag.bst}
\bibliography{REF}

\clearpage
\noindent \textbf{Methods}\\
\noindent \textbf{Fabrication of Si$_3$N$_4$ microresonator}\\
\noindent The Si$_3$N$_4$ microresonator is fabricated in a commercial CMOS foundry. The starting material consists of a 200~mm diameter silicon wafer featuring 14.5 $\upmu$m of thermally-oxidized SiO$_2$ on its surface. The waveguide is formed by low-pressure chemical vapor deposition (LPCVD) of Si$_3$N$_4$, followed by deep-UV lithography and plasma etching for pattern transfer. The resultant Si$_3$N$_4$ core is then capped by 4 $\upmu$m of SiO$_2$, formed using LPCVD with a tetraethyl orthosilicate (TEOS) precursor and annealing at 1150 $^\circ$C. The resultant wafer is singulated into chips using a blade dicing saw.\\

\noindent \textbf{Generation of the self-injection locked microcomb}\\
\noindent We mount a DFB laser chip with an output power of 19 dBm in a butterfly package on a 6-axis stage (Thorlabs MAX602D/M). The Si$_3$N$_4$ chip is placed on a TEC holder, whose temperature is controlled by a temperature controller (Vescent SLICE-QTC) to keep the Si$_3$N$_4$ chip in a stable state. We use a laser diode controller (Thorlabs ITC4001) to adjust the temperature of the DFB and provide the drive current. By using a piezo controller (Thorlabs MDT693B) to control the 6-axis stage, the relative position between the DFB and the Si$_3$N$_4$ chip can be accurately controlled so that the two chips can be closely butt-coupled. By setting the temperature and current of the DFB to the appropriate values, the DFB laser can be self-injection locked, leading to the generation of a microcomb at the through port of the microresonator\cite{shen2020integrated}. The generation of the microcomb can be realized by simply turning on the drive current of the DFB, which is known as the turnkey operation. A lensed fiber fixed on a 3-axis stage (Thorlabs MAX312D/M) is used to collect the output light from the Si$_3$N$_4$ chip. We record the generated microcomb with an optical spectrum analyzer (Yokogawa AQ6370D). By controlling the OSA to periodically store the spectra at 20-second intervals, we characterize the long-term stability of the microcomb over 2 hours.\\

\noindent \textbf{Characterization of the microcomb before and after amplification}\\
\noindent We use a tunable fiber Bragg grating filter (AOS Tunable FBG) to filter one of the comb lines. After adjusting the polarization of the light through a PC (Thorlabs FPC560), we inject it into a DFB laser through a polarization maintaining circulator (PMC) to achieve injection locking amplification. Given the limited number of DFB chips, we adjust the temperature of each DFB laser to modify the output wavelength, enabling the system to achieve amplification of each comb line. 
The PM (Thorlabs PM100D) and the OSA are employed to capture the power and spectral variations before and after injection locking, respectively. 
By observing the change in the output frequency in the OSA while adjusting the temperature of the DFB chip, we can ascertain whether the DFB laser is injection locked.
To measure the frequency noise before and after injection locking, the correlated self-heterodyne method is adopted\cite{yuan2022correlated}. 
We firstly split the light into two paths: one path passes through an acousto-optic modulator (AOM, Gooch\&Housego T-M080-0.4C2J-3-F2P) with a frequency shift of 80 MHz, while the other is delayed by a 5-km-long fiber with a polarization controller. The light is then recombined and split into two beams of equal power received by two identical PDs (Newport 1811-FC). The outputs of PDs are recorded using a high-speed oscilloscope (Keysight MXR404A) and analyzed by a computer. We compare and plot the frequency noise generated using different amplification methods: an EDFA (Amonics AEDFA-C-DWDM), an SOA (Thorlabs SOA1117S) and a DFB laser. Considering that sufficient optical power is needed to make the PD respond efficiently, we measure the linewidth of 21 comb lines of the microcomb. By comparing the linewidth before and after injection locking amplification, we observe that minimal noise is introduced.\\

\noindent \textbf{Coherent communication details}\\
\noindent After being filtered in turn, each comb line passes through a PC and is split into two paths using a polarization maintaining power splitter. One path, with 1\% of the power, connects to a PM for monitoring the polarization of light to ensure effective injection locking. Another path, carrying 99\% of the power, passes through a PMC to IL the DFB.
A very small fraction of the output light from the injection locked DFB is split off and monitored by the OSA. The remaining light is modulated with the data generated by an arbitrary waveform generator (AWG, Keysight M8194A). The modulated signal light and the unmodulated LO are transmitted concurrently through two distinct cores of the 7-core fiber. The LO at the receiver side is amplified to 20 dBm by injection locking another DFB with a similar structure, and it is then sent to the coherent receiver together with the signal light. The receiver is powered by a DC power supply (Keysight E36312A). The electrical signals generated by the receiver through the beating of the signal light with the LO are captured by a high-performance real-time oscilloscope after passing through DC blocks (Gwave GDCB-67G-185) and baluns (HYPERLABS HL9407), respectively. The OSC can provide a conversion rate of 80 GSa/s and a bandwidth of 33 GHz on four simultaneous channels. The data samples we collect at a time last for 6.5 $\upmu$s and are subsequently processed by a computer for DSP.\\

\noindent \textbf{Digital signal processing}\\
\noindent In the DSP at the receiver side, the sampled signal is first orthogonally normalized using the Gram-Schmidt algorithm to compensate for the effects of non-ideal factors in the link. Continuing with the process, we apply matched filtering to the signal using a root-raised cosine filter with a roll-off coefficient of 0.05, followed by downsampling the signal from 80 GSa/s to 30 GSa/s. Subsequently, adaptive equalization and carrier recovery are performed. The equalizer adopts a $4\times2$ multiple-input multiple-output (MIMO) structure. We employ the constant modulus algorithm (CMA) to achieve pre-convergence of the taps in the equalizer. The output signal from CMA is then utilized for FOE. FOE is accomplished by taking the signal to the fourth power and identifying the highest spectral peak. After this, a decision-directed approach is employed for equalization. In equalization, CPE is executed using the blind phase search (BPS) algorithm with 16 test angles. After equalization and carrier recovery, orthogonalization is employed to address both hybrid imperfections and modulator bias issues. Finally, symbol decisions are made\cite{faruk2017digital}. BERs are calculated to assess the system performance, utilizing over 1,920,000 bits.\\

\noindent\textbf{Design and fabrication of the transceiver}\\
\noindent The transmitter and receiver are both SOI-based SiPh integrated chips. At the transmitter side, the optical carrier is split into two paths that pass through two IQ modulators. To maintain the correct operating state, the modulators are monitored by MPDs and adjusted by heaters. The drivers are used to drive the modulators to load electrical signals onto the optical carrier. Two paths of modulated lights go into the PRBC to give a dual-polarized IQ-modulated signal, which is then coupled into the optical fiber through a fiber array (FA) for transmission. At the receiver side, The data carrier and LO are coupled to the receiver chip via an FA. The PRBS splits the data carrier into two polarization-orthogonal paths. The power of each path is monitored by MPDs and adjusted by VOAs. The signal lights and LOs are mixed in $90^\circ$ optical hybrids and converted to photocurrent by BPDs. The current is converted to voltage output and amplified by TIAs.\\

\noindent\textbf{Data availability}\\
\noindent The data that supports the plots within this paper and other findings of this study are available from the corresponding authors upon reasonable request.\\ 

\noindent\textbf{Code availability}\\
\noindent The codes that support the findings of this study are available from the corresponding authors upon reasonable request.\\

\vspace{12pt}
\begin{footnotesize}

\vspace{6pt}
\noindent \textbf{Acknowledgments}\\
\noindent We thank C. Zhang, X. Chen and M. Zuo for helpful discussions. The experiments are supported by High-performance Computing Platform of Peking University.

\vspace{6pt}
\noindent \textbf{Author contributions}\\
\noindent 
The concept of this work was conceived by X.Z., L.C. and X.W. The experiments were performed by X.Z., Z.Z. and Y.G., with the assistance of M.Z., W.J., B.S., H.S., Y.C., J.H., Z.T., M.J., R.C., Z.G., Z.F., N.Z., Y.L., P.C., W.H. and D.P. The results were analyzed by X.Z., Z.Z., Y.G. and M.Z. All authors participated in the writing of the manuscript. The project was under the supervision of L.C., X.W. and J.E.B.

\vspace{6pt}
\noindent \textbf{Competing financial interests} \\
\noindent The authors declare no competing financial interests.

\vspace{6pt}
\noindent \textbf{Additional information} \\
\noindent \textbf{Supplementary information} is available for this paper.\\
\noindent \textbf{Correspondence and requests for materials} should be addressed to Lin Chang, Xingjun Wang or John E. Bowers.\\
\noindent \textbf{Peer review information} \\
\noindent \textbf{Reprints and permissions information} is available at www.nature.com/reprints.\\

\end{footnotesize}

\end{document}


\noindent
{\large Supplementary Information for}\\
\title{High-coherence parallelization in integrated photonics}
\author{Xuguang Zhang$^{1,\dagger}$, ZiXuan Zhou$^{1,\dagger}$, Yijun Guo$^{1,\dagger}$, Minxue Zhuang$^{1}$, Warren Jin$^{2}$, Bitao Shen$^{1}$, Yujun Chen$^{1}$, Jiahui Huang$^{1}$, Zihan Tao$^{1}$, Ming Jin$^{1}$, Ruixuan Chen$^{1}$, Zhangfeng Ge$^{5}$, Zhou Fang$^{4}$, Ning Zhang$^{4}$, Yadong Liu$^{4}$, Pengfei Cai$^{4}$, Weiwei Hu$^{1}$, Haowen Shu$^{1}$, Dong Pan$^{4}$, John E. Bowers$^{2,*}$, Xingjun Wang$^{1,3,5,*}$ \& Lin Chang$^{1,3,*}$\\
\vspace{3pt}
$^1$State Key Laboratory of Advanced Optical Communications System and Networks, School of Electronics, Peking University, Beijing, China.\\
$^2$Department of Electrical and Computer 
Engineering, University of California Santa Barbara, Santa Barbara, CA, USA.\\
$^3$Frontiers Science Center for Nano-optoelectronics, Peking University, Beijing, China.\\
$^4$SiFotonics Technologies Co., Ltd., Beijing, China.\\
$^5$Peking University Yangtze Delta Institute of Optoelectronics, Nantong, China.\\
$^\dagger$These authors contributed equally to this work. \\
\vspace{3pt}
\centering{\small Corresponding authors: $^*$bowers@ece.ucsb.edu, $^*$xjwang@pku.edu.cn, $^*$linchang@pku.edu.cn}}

\maketitle
\clearpage



\begin{figure*}[htp]
    \vspace{-1 pc}
    \centering
    \includegraphics[width=0.7\linewidth]{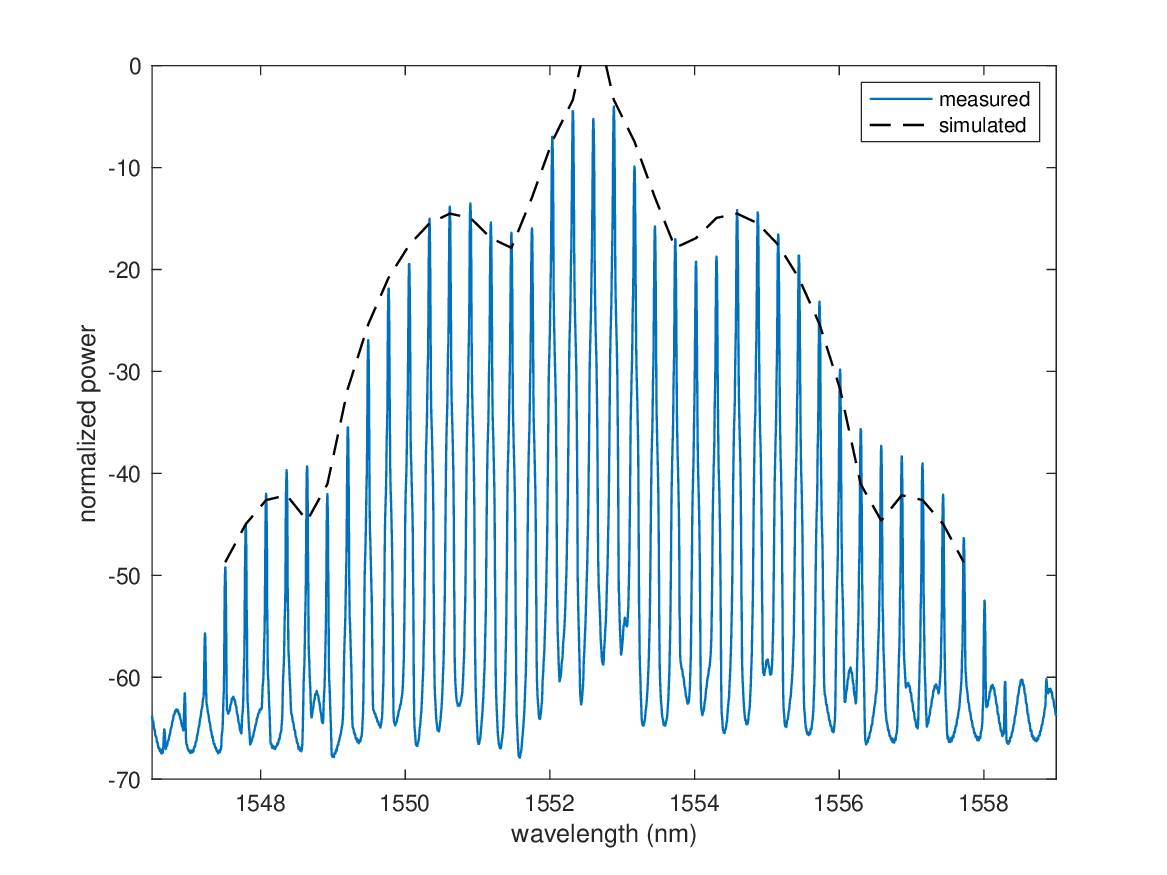}
    \caption{Simulated and measured microcomb spectra}
    \label{figSI_4.1}
\end{figure*}

\noindent \textbf{Supplementary note \uppercase\expandafter{\romannumeral 1}: The spectrum of Kerr comb in the dark pulse regime}\\ 
\noindent The Si$_3$N$_4$ microresonator used in our experiment works in normal group velocity dispersion (GVD), resulting in the generation of the dark-pulse microcomb. The microcomb state in a microresonator can be simulated by solving the Lugiato-Lefever equation (LLE)\cite{godey2014stability, chembo2013spatiotemporal}:

\begin{equation}
\begin{aligned}
\frac{\partial \psi}{\partial \tau} = -(1+i\alpha)\psi + i|\psi|^2 \psi 
- i\frac{\beta}{2} \frac{\partial^2 \psi}{\partial \theta^2} + F
\end{aligned}
\end{equation}

\noindent in which $\psi(\theta,\tau)$ denotes the normalized intracavity electric field envelope, while $\alpha$, $\beta$ and $F$ are normalized pump-cavity detuning, second-order microresonator dispersion and pump electric field, respectively. Using the parameters $\alpha=6$, $\beta=0.12$, and $F=2.92$ with an initial condition of $\psi (\theta,0)=2-1.8 e^{-(\theta/0.2)^2}$, we compare the simulated and measured comb spectra, as shown in Fig. \ref{figSI_4.1}. This indicates that the comb generation process in our experiment is consistent with the theory.\\

\indent The normalized parameters represent a pump-cavity detuning of $\Delta \Omega=-\frac{\omega_p \alpha}{2Q}=-76.5\text{ MHz}$ and a second-order microresonator dispersion of $D_2=-\frac{\omega_p \beta}{2Q}=-1.53\text{ MHz}$.\\

\newpage
\noindent \textbf{Supplementary note \uppercase\expandafter{\romannumeral 2}: Principle of the DFB Laser Injection Locking}\\
\noindent The equation for the electric field $E$ (normalized so that photon number $S(t)=E(t)^2$) in the slave DFB laser is\cite{lang1982injection}:

\begin{equation}
\begin{aligned}
	\frac{dE(t)}{dt}=\frac12 (1 + i\alpha)\Big(g(N(t)-N_{tr})-\gamma_p \Big)E(t) - i(\omega_{m}-\omega_{s,fr})E(t) + \kappa A_{inj} e^{i\phi_{inj}}
\end{aligned}
\end{equation}

\noindent where $g$, $N_{tr}$, $\gamma_p$, $\omega_m$, $\omega_{fr}$, $\alpha$, $\kappa$, $A_{inj}$ and $\phi_{inj}$ denote linear gain coefficient, transparent carrier number, photon decay rate, master (seed) light frequency, free-running frequency of slave laser, semiconductor linewidth enhancement factor, face-let coupling rate, master light electric field amplitude, and master light electric field phase respectively. 
The three items on the right hand side represent gain, injection light frequency detuning and injected light eclectic field.\\
\indent By decomposing the slave laser electric field $E$ into intensity (intracavity photon number $S$) and phase $\phi$, and combining the equation of the slave laser carrier number $N$, we get the basic IL equations\cite{lau2009enhanced, liu2020optical}: 

\begin{align}
	\frac{dS(t)}{dt} &= \Big(g(N(t)-N_{tr})-\gamma_p \Big)S(t) + 2\kappa \sqrt{S_{inj} S(t)} \cos (\phi(t)-\phi_{inj})
	\\
	\frac{d\phi (t)}{dt} &= \frac{\alpha}{2} \Big(g(N(t)-N_{tr})-\gamma_p \Big) - (\omega_{m}-\omega_{s,fr}) - \kappa \sqrt{\frac{S_{inj}}{S(t)}}\sin(\phi(t)-\phi_{inj})
    \\
    \frac{dN(t)}{dt} &= J(t) - \gamma_N N(t) - g\Big(N(t)-N_{tr}\Big)S(t)
\end{align}

\noindent where $J(t)$ denotes the DFB laser current and $\gamma_N$ represents carrier recombination rate. 
Parameter meters used in our calculation are listed in 
Table \ref{injection locked parameters}.\\

\begin{figure*}[htp]
    \vspace{-1 pc}
    \centering
    \includegraphics[width=0.7\linewidth]{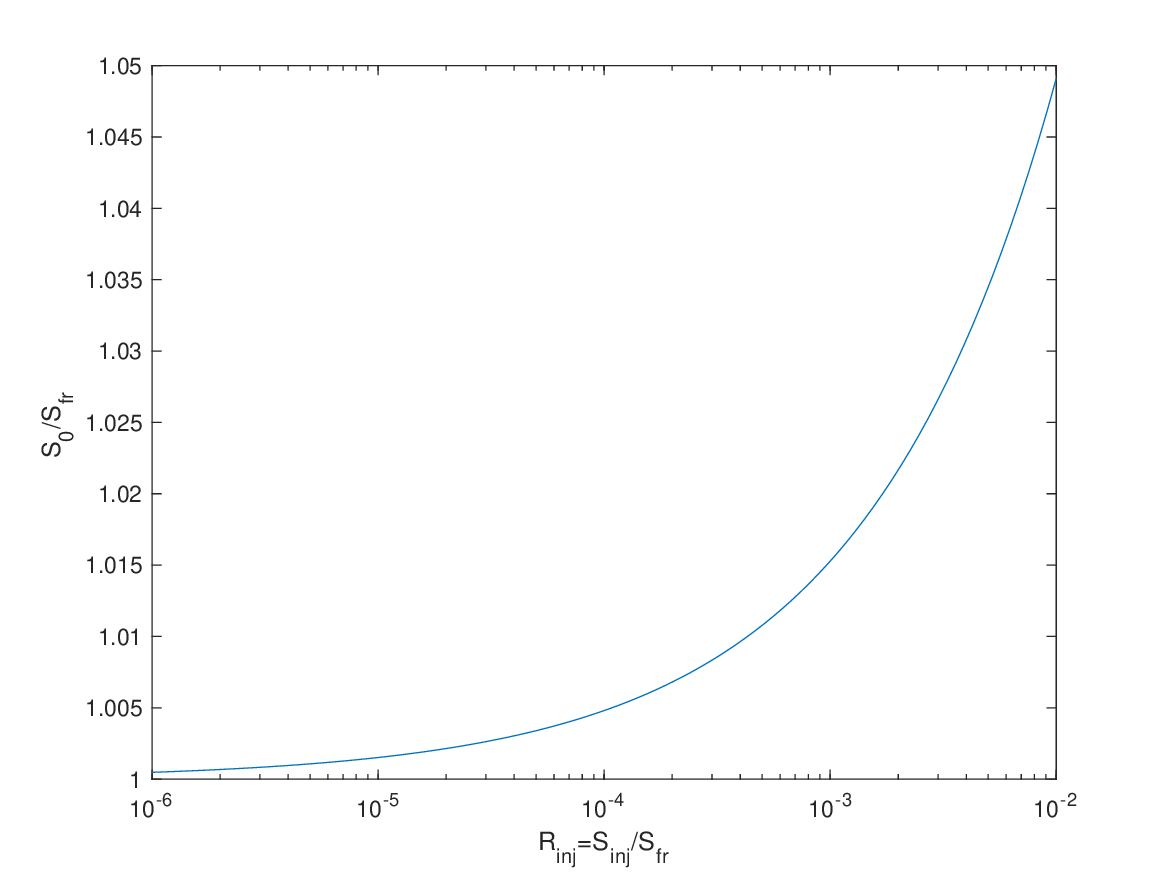}
    \caption{Amplification effect in injection locked laser}
    \label{figSI_5.1}
\end{figure*}

\begin{table}[htbp]
\centering
\setlength{\tabcolsep}{19mm}{
\begin{tabular}{ccc}
\hline
\textbf{Parameter} & \textbf{Value} & \textbf{Unit} \\ \hline
$g$                & 4.7e4          & $s^{-1}$      \\
$N_{tr}$           & 2.7e+08        &               \\
$\alpha$           & 5              &               \\
J                  & 1.53e18        & $s^{-1}$      \\
$\gamma_N$         & 1e9            & $s^{-1}$      \\
$\gamma_P$         & 100e9          & $s^{-1}$      \\
$\kappa$           & 2.4e10         & $s^{-1}$      \\
$\Delta \phi_0$    & 1.3e-1         &               \\ \hline
\end{tabular}}
\caption{Injection Locking Parameters}
\label{injection locked parameters}
\end{table}
\noindent By solving the steady rate solution in which 
\begin{align}
	\frac{dS(t)}{dt} &= 0 \\
	\frac{d\phi(t)}{dt} &= 0 \\
	\frac{dN(t)}{dt} &= 0
\end{align}
we obtain that:
\begin{align}
	S_0 &= \frac{ S_{fr}- \frac{\gamma_N}{\gamma_p} \Delta N_0}{1+\frac{g}{\gamma_p} \Delta N_0} \\ 
	\Delta \phi_0 &= \phi_0 - \phi_{inj} = \sin^{-1} \Big(-\sqrt{\frac{S_0}{S_{inj}}} \frac{\omega_m-\omega_{s,fr}}{\kappa \sqrt{1+\alpha^2}}\Big)-\tan^{-1}\alpha \\
	\Delta N_0 &\equiv N_0 - N_{tr} - \frac{\gamma_p}{g} = -\frac{2\kappa}{g} \sqrt{\frac{S_{inj}}{S_0}} \cos \phi_0 
\end{align}
\indent From the solution above, we can analyze the relationship between input light power and output light power, which is denoted by injected light photon number $S_{inj}$ and intracavity photon number $S_{0}$. Since injection light and output light go through the same facet, the ratio of $S_{0}$ to $S_{inj}$ is the amplification gain. The relationship between $S_{0}$ and $S_{inj}$ (normalized by free-running intracavity number $S_{fr}$) is depicted in Fig \ref{figSI_5.1}. In our amplification case where $\frac{S_{inj}}{S} << 1$ and $\Delta N << 1$, the output light power is pretty close to the free-running power of the slave laser $S \sim S_{fr}$ with a difference up to 5\% over a width range of injection power ratio from -60 dB to -20 dB. 

\begin{figure*}[htbp]
    \centering
    \includegraphics[width=1\linewidth]{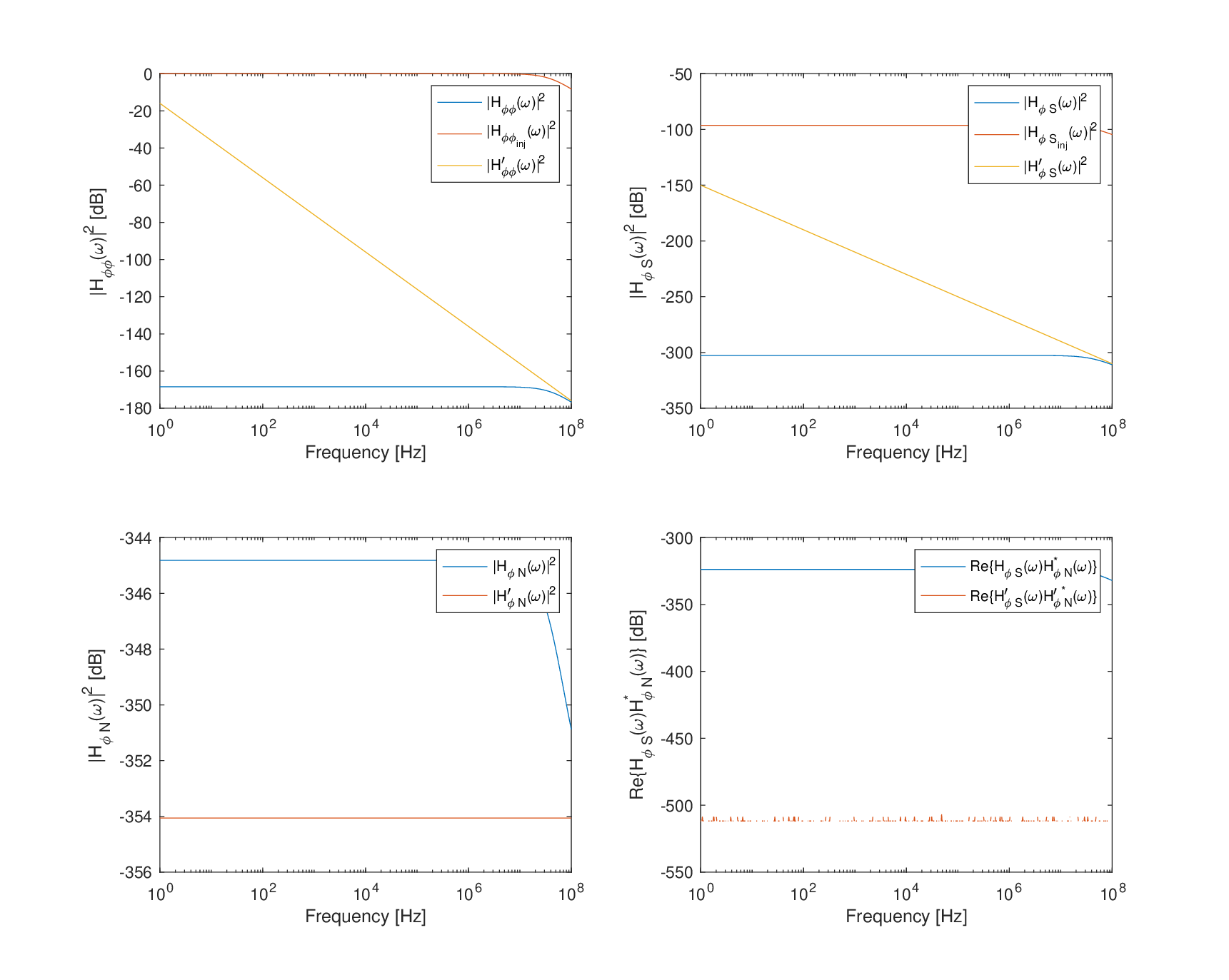}
    \caption{Injection locked output light phase noise decomposition}
    \label{figSI_5.2}
\end{figure*}

\begin{figure*}[htbp]
    \centering
    \includegraphics[width=0.6\linewidth]{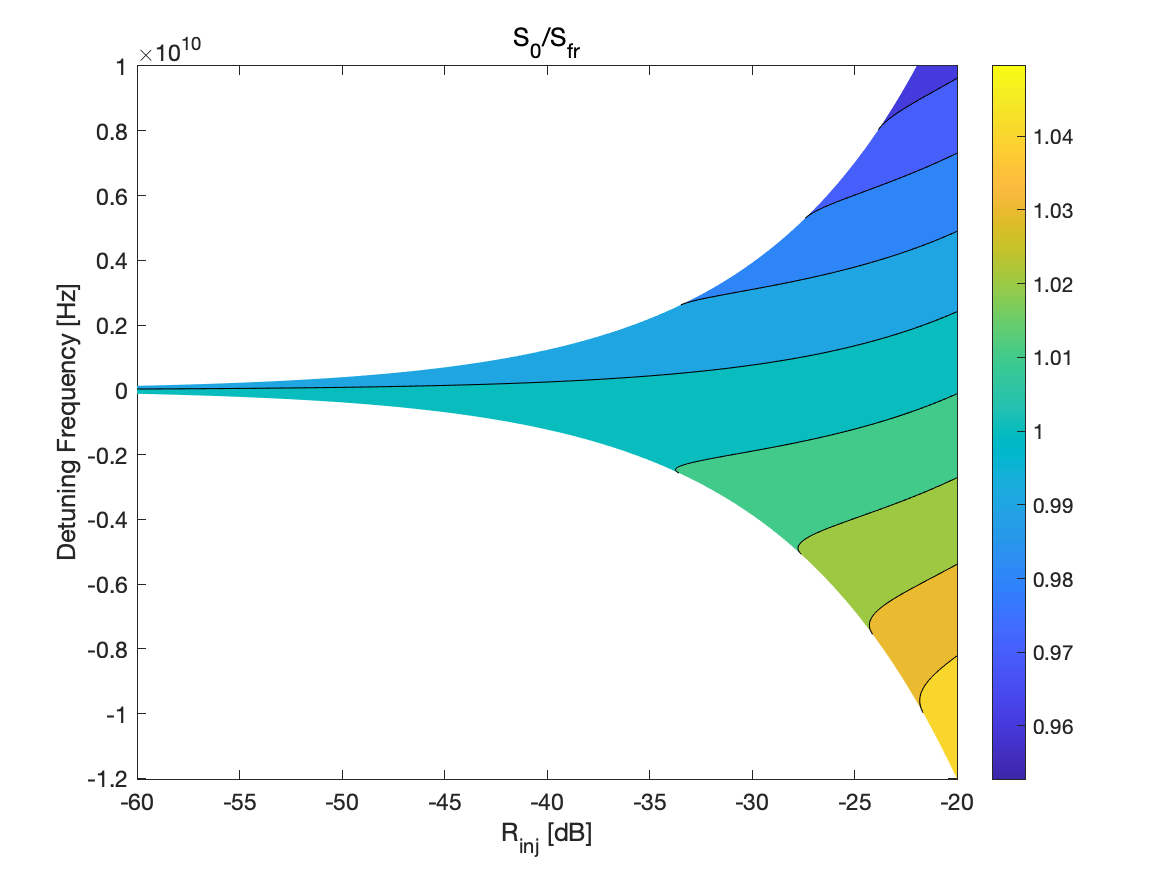}
    \caption{Injection locking bandwidth versus injection ratio}
    \label{figSI_5.3}
\end{figure*}

\noindent We consider small perturbations near the steady state solution to analyze the phase noise in injection locked DFB: 
\begin{align}
	S(t) &= S_0 + \delta S(t) \\
	\phi (t) &= \Delta\phi_0 + \phi_{inj} + \delta \phi(t) \\
	\Delta N(t) &= \Delta N_0 + \delta N(t)
\end{align}
\noindent By expanding the basic IL equations to the first order, we obtain the matrix form of the IL small signal solution:
\begin{equation}
\begin{aligned}
	\begin{bmatrix}
	m_{SS}+s & m_{S\phi} & m_{SN} \\
	m_{\phi S} & m_{\phi\phi}+s & m_{\phi N} \\
	m_{NS} & 0 & m_{NN}+s
	\end{bmatrix}
	&\begin{bmatrix}
	\delta \tilde{S} \\
	\delta \tilde{\phi} \\
	\delta \tilde{N}
	\end{bmatrix} \\
	= 
	\begin{bmatrix}
	0 \\ 0 \\ \delta \tilde{J}
	\end{bmatrix}
	+&\begin{bmatrix}
	m_{Si} \\
	m_{\phi i} \\
	0
	\end{bmatrix} \delta \tilde{S}_{inj}
	+\begin{bmatrix}
	m_{S\phi} \\
	m_{\phi\phi}\\
	0
	\end{bmatrix} \delta \tilde{\phi}_{inj}
	+\begin{bmatrix}
	\tilde{F}_S \\
	\tilde{F}_\phi\\
	\tilde{F}_N
	\end{bmatrix}
\end{aligned}
\end{equation}
\noindent where $F_S$, $F_\phi$ and $F_N$ denote the s-domain photon, phase, and carrier Langevin noise sources, and the parameters in which are
\begin{align}
	m_{SS} &= -g\Delta N_0 - \kappa \sqrt{\frac{S_{inj}}{S_0}}\cos\Delta\phi_0 = \kappa \sqrt{\frac{S_{inj}}{S_0}}\cos\Delta\phi_0\\
	m_{S\phi} &= 2\kappa \sqrt{S_{inj}S_0} \sin\Delta\phi_0 \\
	m_{SN} &= -g S_0 \\
	m_{Si} &= \kappa \sqrt{\frac{S_{inj}}{S_0}}\cos\phi_0\\
	m_{\phi S} &= -\frac{\kappa}{2} \sqrt{\frac{S_{inj}}{S_0^3}} \sin \Delta\phi_0 \\
	m_{\phi \phi} &= \kappa \sqrt{\frac{S_{inj}}{S_0}}\cos\Delta\phi_0 \\
	m_{\phi N} &= -\frac{g}{2} \alpha \\
	m_{\phi i} &= -\frac{\kappa}{2} \sqrt{\frac{1}{S_0 S_{inj}}}\sin\Delta\phi_0\\
	m_{NS} &= \gamma_p + g\Delta N_0 = \gamma_p - 2\kappa \sqrt{\frac{S_{inj}}{S_0}}\cos\Delta\phi_0 \\
	m_{NN} &= \gamma_N + g S_0.
\end{align}
\noindent Since the process is stationary, we can derive the power spectral density of the phase noise:
\begin{equation}
\begin{aligned}
\begin{split}\\
	<\delta \phi \delta \phi>_\omega =& |H_{\phi S_i}(\omega)|^2<\delta S_{inj}\delta S_{inj}>_\omega + |H_{\phi \phi_i}(\omega)|^2<\delta \phi_{inj}\delta \phi_{inj}>_\omega \\
	&+ |H_{\phi S}(\omega)|^2 <F_S F_S>_\omega + |H_{\phi \phi}(\omega)|^2 <F_\phi F_\phi>_\omega + |H_{\phi N}(\omega)|^2 <F_N F_N>_\omega \\
	&+ 2Re\{H_{\phi S}(\omega) H_{\phi N}^*(\omega)\}<F_S F_N>_\omega
\end{split}
\end{aligned}
\end{equation}
\noindent where
\begin{align}
    \begin{split}
	H_{\phi S_i}(s) =& \Big[m_{Si}\Big(m_{\phi N}m_{SA} - m_{\phi S}(m_{NN}+s)\Big) \\
		&+ m_{\phi i}\Big((m_{SS}+s)(m_{NN}+s)-m_{SN}m_{NS}\Big)\Big]/D(s)
    \end{split}\\
    \begin{split}
	H_{\phi\phi_i}(s) =& \Big[m_{S\phi}\Big(m_{\phi N}m_{NS}-m_{\phi S}(m_{NN}+s)\Big) \\
		&+ m_{\phi \phi}\Big((m_{SS}+s)(m_{NN}+s)-m_{SN}m_{NS}\Big)\Big]/D(s)
    \end{split}\\
	H_{\phi S}(s) =& \Big(m_{\phi N}m_{NS} - m_{\phi S}(m_{NN}+s)\Big)/D(s)\\
	H_{\phi\phi}(s) =& \Big((m_{SS}+s)(m_{NN}+s) - m_{NS}m_{SN}\Big)/D(s) \\
	H_{\phi N}(s) =& \Big(m_{\phi S}m_{SN} - m_{\phi N}(m_{SS}+s)\Big)/D(s) \\
	D(s) =& \det
		\begin{bmatrix}
			m_{SS}+s & m_{S\phi} & m_{SN} \\
			m_{\phi S} & m_{\phi\phi}+s & m_{\phi N} \\
			m_{NS} & 0 & m_{NN}+s
		\end{bmatrix}
\end{align}
\noindent In the free-running case, the phase noise is only relevant to the Langevin noise: 
\begin{equation}
\begin{aligned}
	<\delta \phi \delta \phi>_\omega =& 
	~|H_{\phi S}^{\prime}(\omega)|^2 <F_S F_S>_\omega + |H_{\phi \phi}^{\prime}(\omega)|^2 <F_\phi F_\phi>_\omega + |H_{\phi N}^{\prime}(\omega)|^2 <F_N F_N>_\omega \\
	&+ 2Re\{H_{\phi S}^{\prime}(\omega) H_{\phi N}^{\prime *}(\omega)\}<F_S F_N>_\omega
\end{aligned}
\end{equation}
\indent We calculate and compare relevant terms in Fig. \ref{figSI_5.2}.
The frequency response of laser cavity Langevin phase noise $|H_{\phi,\phi}(\omega)|^2$ can be suppressed by more than 40 dB compared with the free-running case $|H_{\phi,\phi}(\omega)^{\prime}|^2$, while that of the master light $|H_{\phi,\phi_{inj}}|$ remains close to 1 in the frequency range of our interest (Fig. \ref{figSI_5.2}, top-left). 
Injection locking process also greatly suppresses the part of phase noise due to laser cavity Langevin intensity noise (Fig. \ref{figSI_5.2}, top-right), while the carrier number fluctuation and carrier-photon coupling fluctuation have a relatively minor impact on total phase noise (Fig. \ref{figSI_5.2}, down). In this way, we demonstrate that the injection locking technique has a great potential in suppressing the original DFB phase noise. \\
\indent The limitation of our injection locking amplification gain may lie in the locking bandwidth. 
By analyzing the steady rate solution $\Big| \sqrt{\frac{S_0}{S_{inj}}} \frac{\omega_m-\omega_{s,fr}}{\kappa \sqrt{1+\alpha^2}} \Big| < 1$ and the stability of small signal solution (the zeros of $D(s)$ lie in the left half of the s-plane), we can calculate the bandwidth in which injection locking is stable (Fig. \ref{figSI_5.3}).
Under the injection ratio $R_{inj}=S_{inj}/S_{fr}$ of -60 dB, the locking bandwidth is narrowed down to nearly 200 MHz. Under this circumstance, the locking state is vulnerable to thermal and mechanical resonance frequency drift. 
The amplification ability can be further enhanced by designing a semiconductor laser with a higher coupling coefficient $\kappa$ (lower cavity quality factor), a higher linewidth enhancement factor $\alpha$, or applying an optical phase lock loop (OPLL) technique.\\



\begin{figure*}[htbp]
    \centering
    \includegraphics[width=0.7\linewidth]{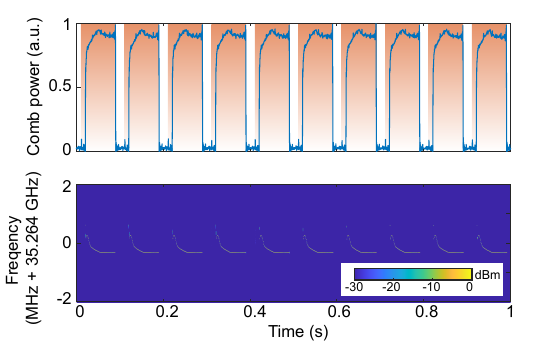}
    \caption{Measured comb power and spectrogram of
comb repetition rate versus time.}
    \label{turnkey}
\end{figure*}

\newpage
\noindent \textbf{Supplementary note \uppercase\expandafter{\romannumeral 3}: Turnkey microcomb generation}\\ 
\noindent We apply a predefined periodic square wave current to the DFB chip to simulate turnkey operation. The microcomb with 1-FSR spacing can be rapidly and stably established in each period. We utilize a tunable FBG filter to suppress the pump light and employ a photodetector to obtain the repetition rate signal in Fig. \ref{turnkey}.\\

\begin{figure*}[htbp]
    \centering
    \includegraphics[width=0.6\linewidth]{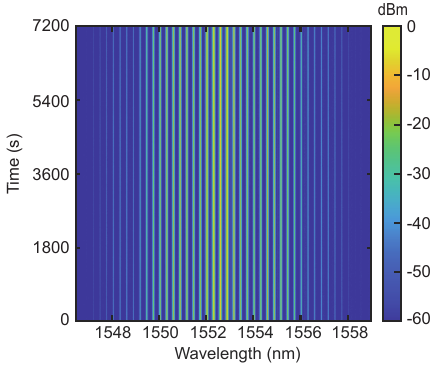}
    \caption{The spectra of the microcomb over time}
    \label{long time}
\end{figure*}

\noindent \textbf{Supplementary note \uppercase\expandafter{\romannumeral 4}: Long-term stability of the self-injection locked microcomb}\\ 
\noindent After generating the microcomb by self-injection locking the DFB to the Si$_3$N$_4$ microresonator, we record the spectra using an optical spectrum analyzer (OSA) every 20 seconds for 2 hours in Fig. \ref{long time}. The microcomb demonstrates high stability during this timeframe, with power fluctuation of less than 2\%, possibly attributed to random mechanical vibrations and temperature perturbations in the external environment.\\

\begin{figure*}[htbp]
    \centering
    \includegraphics[width=0.5\linewidth]{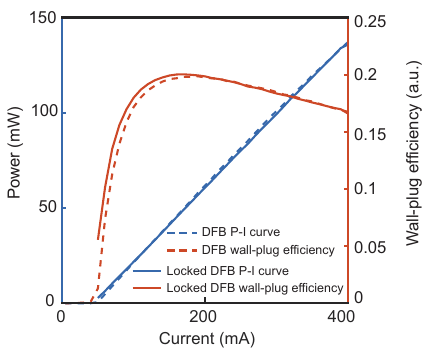}
    \caption{P-I curve and WPE of the DFB before and after locking}
    \label{WPE}
\end{figure*}

\noindent \textbf{Supplementary note \uppercase\expandafter{\romannumeral 5}: The characteristics of the DFB before and after injection locking}\\ 
\noindent We select one of the comb lines using a filter, attenuate the power to below -20 dBm with a variable optical attenuator (VOA), and inject the light into the DFB through a circulator for locking. We read out the voltage applied to the DFB chip from the panel of the controller which drives current on the DFB. Simultaneously, we measure the output optical power from the circulator using an optical power meter (PM) and calculate the corresponding wall-plug efficiency (WPE) in Fig. \ref{WPE}. From the measured P-I curves and the WPE, it can be noted that the output optical power and WPE of the DFB is nearly unaffected after being injection locked by a comb line. This is consistent with the theoretical analysis.\\

\begin{figure*}[htbp]
    \centering
    \includegraphics[width=\linewidth]{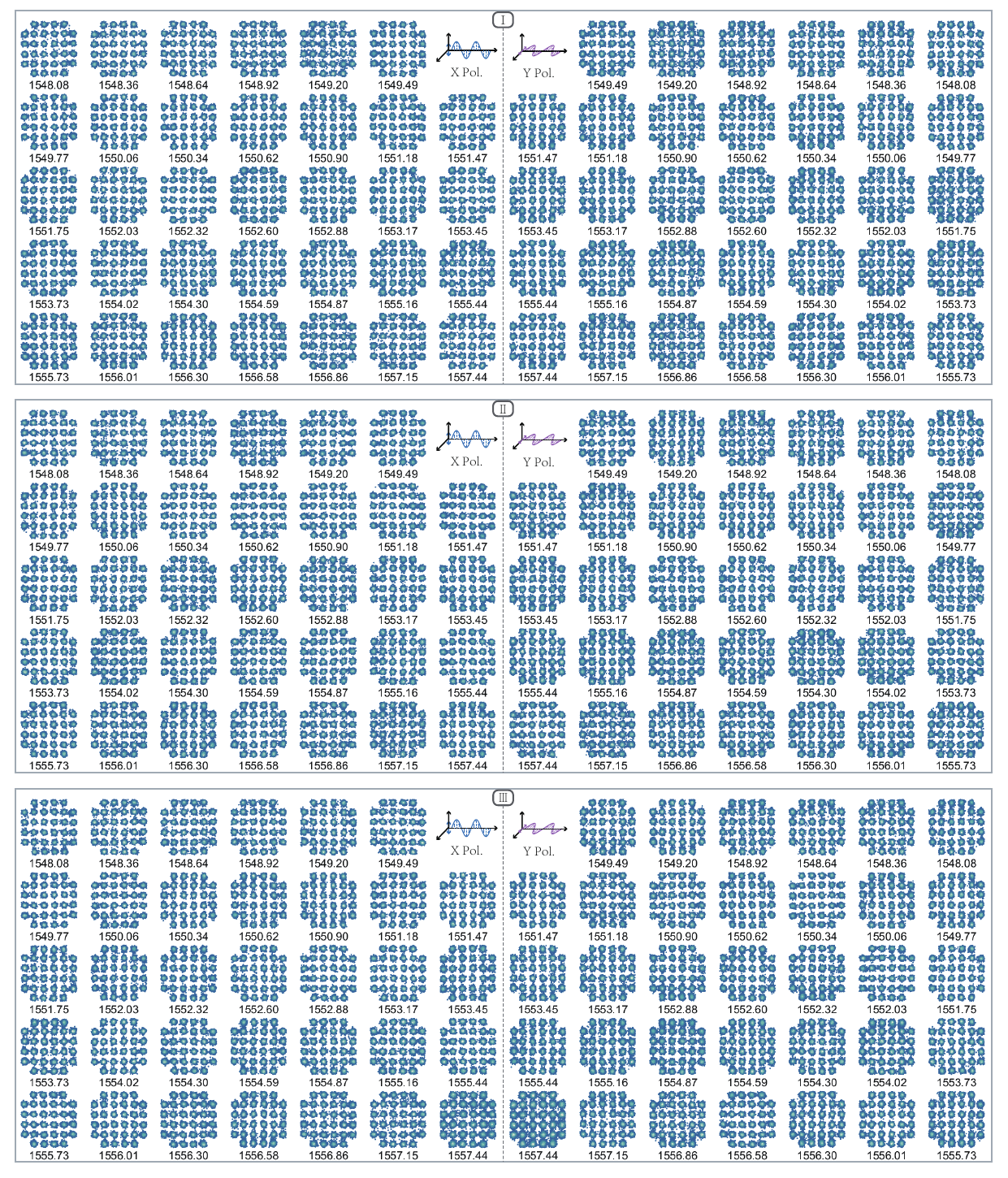}
    \caption{The recovered constellation diagrams of all 34 data channels in core 1, 2 and 3}
    \label{core1-3}
\end{figure*}

\begin{figure*}[htbp]
    \centering
    \includegraphics[width=\linewidth]{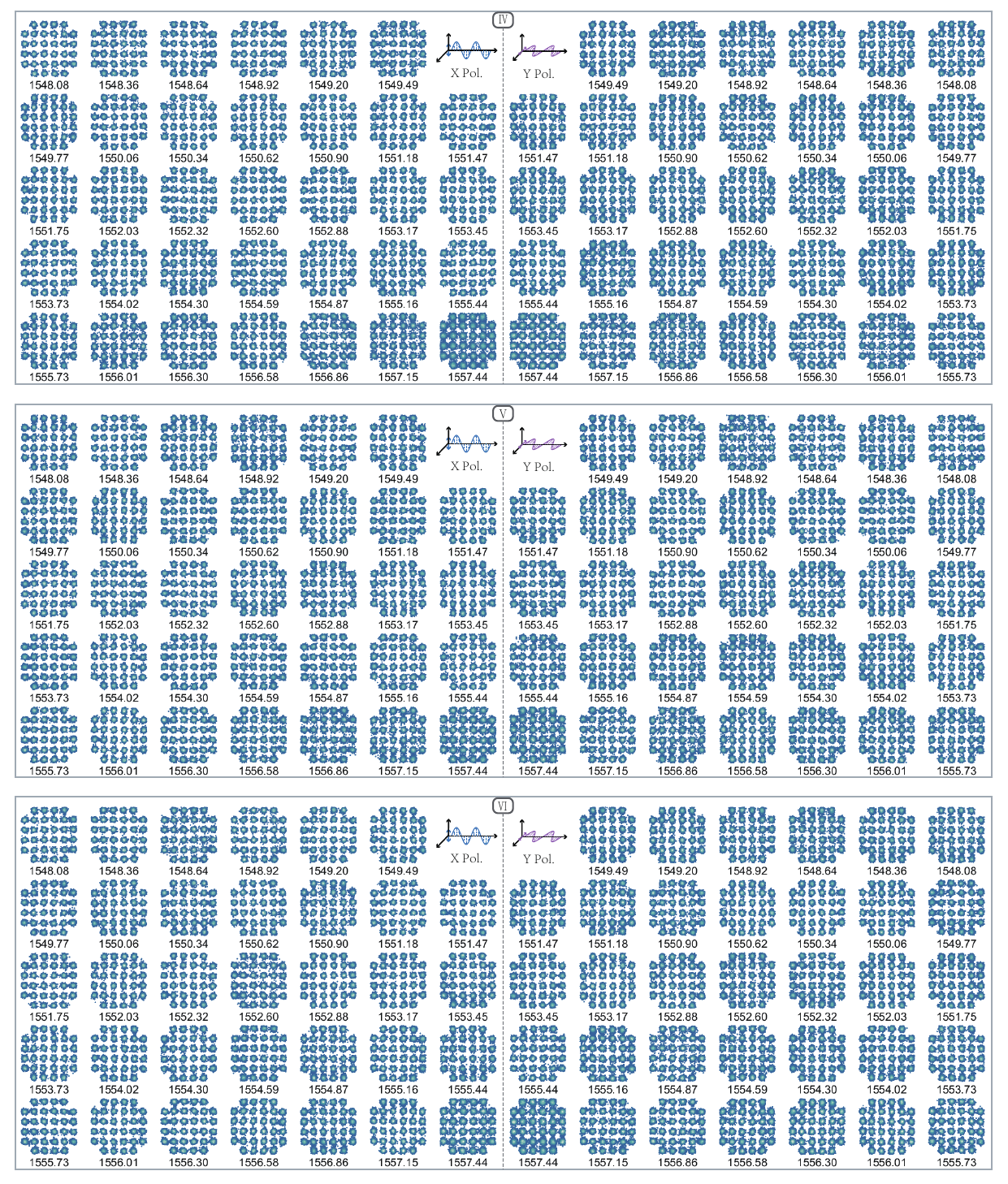}
    \caption{The recovered constellation diagrams of all 34 data channels in core 4, 5 and 6}
    \label{core4-6}
\end{figure*}

\newpage
\noindent \textbf{Supplementary note 
\uppercase\expandafter{\romannumeral 6}: Results of the 7-core fiber WDM communication}\\ 
\noindent We perform the 7-core fiber WDM communication experiment described in the main text. We use 6 of the cores for data transmission and successfully recover all constellation diagrams, as depicted in Fig. \ref{core1-3} and Fig. \ref{core4-6}. 

As shown in the figure, all wavelength channels in signal cores (1-6) show uniform and stable transmission results, whose BERs are all much lower than the 7\% hard-decision forward error correction (HD-FEC) threshold of $3.8\times10^{-3}$.\\

\begin{figure*}[htbp]
    \centering
    \includegraphics[width=0.95\linewidth]{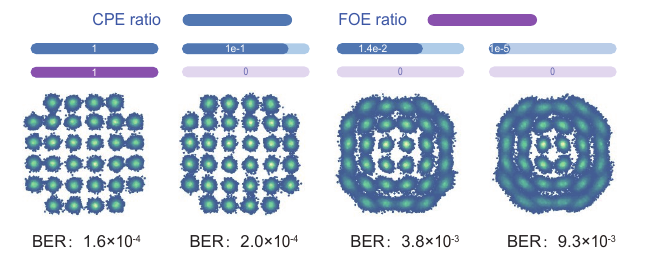}
    \caption{Constellation diagrams with different ratios of CPE and FOE when using DFB locked by a free-running DFB}
    \label{QAM}
\end{figure*}

\newpage
\noindent \textbf{Supplementary note \uppercase\expandafter{\romannumeral 7}: Communication result using DFB locked by a free-running DFB}\\ 
\noindent We use a free-running DFB laser as the source which is split into two parts. One as a data carrier carries 99\% of the power, and the other with only 1\% power serves as the LO. The data carrier and the LO are transmitted concurrently. At the receiver side, the LO is amplified by injection locking another DFB. We perform an experimental transmission of a 30 GBaud DP-32-QAM signal.\\
\indent In this scheme, the FOE can be omitted. However, as we gradually decrease the CPE ratio, the results deteriorate (Fig. \ref{QAM}). When we reduce the CPE ratio to $1.4\times10^{-2}$, the BER increases to  $3.8\times10^{-3}$, the threshold of 7\% HD-FEC. Continuing to decrease the CPE ratio to $1\times10^{-5}$ results in a further increase in the BER, reaching $9.3\times10^{-3}$.\\

\begin{figure*}[htbp]
    \centering
    \includegraphics[width=0.5\linewidth]{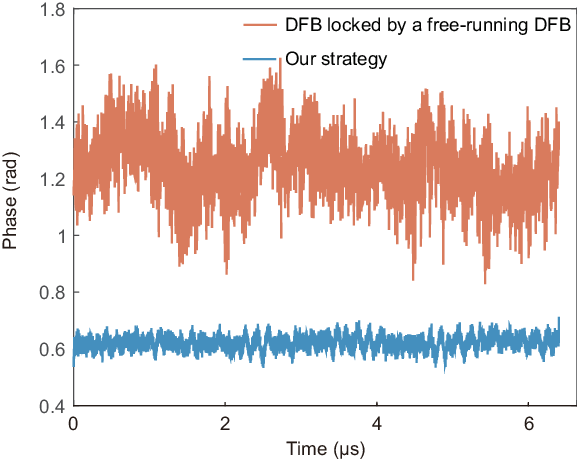}
    \caption{Phase fluctuations for different schemes}
    \label{phase}
\end{figure*}

\indent The phase fluctuation of the DFB locked by a free-running DFB is shown in Fig. \ref{phase}. It can be seen that the DFB locked by a free-running DFB exhibits a larger phase fluctuation compared to our approach. Consequently, decreasing the proportion of CPE markedly increases the BER in this scenario. However, in the context of our strategy, the BER does not deteriorate.\\

\bibliographystyle{naturemag.bst}
\bibliography{REF}